\documentclass[conference]{IEEEtran}[10pt]

\ifCLASSINFOpdf
   \usepackage[pdftex]{graphicx}
\else
\fi
\usepackage{amsmath,amssymb,amsfonts}
\usepackage{braket}
\usepackage{xspace}
\usepackage[dvipsnames]{xcolor}
\usepackage[nolist]{acronym}
\usepackage{amssymb}
\usepackage{bm}

\usepackage{subfig}
\usepackage{graphicx}
\usepackage{comment}

\usepackage{tikz}
\usetikzlibrary{arrows.meta, positioning}

\usepackage[compact]{titlesec}

\begin{document}
\title{On the Achievable Rate of  Satellite Quantum Communication Channel using Deep Autoencoder Gaussian Mixture Model}

\author{\IEEEauthorblockN{Mouli~Chakraborty \IEEEauthorrefmark{1}, Subhash~Chandra \IEEEauthorrefmark{1}, Avishek~Nag \IEEEauthorrefmark{3},  Anshu~Mukherjee \IEEEauthorrefmark{2}\IEEEauthorrefmark{4}}
\IEEEauthorblockA{\IEEEauthorrefmark{1} School of Natural Sciences,
Trinity College Dublin, The University of Dublin, College Green, Dublin 2, Ireland\\
\IEEEauthorrefmark{3} School of Computer Science,
University College Dublin, Belfield, Dublin 4, Ireland\\
\IEEEauthorrefmark{2} School of Electrical and Electronic Engineering,
University College Dublin, Belfield, Dublin 4, Ireland, \\ \IEEEauthorrefmark{4} Beijing-Dublin International College, Beijing University of Technology, Chaoyang, Beijing, China\\
Email:  moulichakraborty@ieee.org,  schandra@tcd.ie, avishek.nag@ucd.ie, anshu.mukherjee@ieee.org}}

\maketitle

\begin{abstract}

We present a comparative study \ac{GMM} and \ac{DAGMM} for estimating satellite quantum channel capacity, considering \ac{HQN} and transmission constraints. While \ac{GMM} is simple and interpretable, \ac{DAGMM} better captures non-linear variations and noise distributions. Simulations show that \ac{DAGMM} provides tighter capacity bounds and improved clustering. This introduces the \ac{DCGMM} for high-dimensional quantum data analysis in quantum satellite communication.

\end{abstract}

\IEEEpeerreviewmaketitle


\begin{IEEEkeywords}
 \ac{HQN}, Satellite Quantum Channel, Achievable Rate, Poisson Noise, Gaussian Mixture Model, Deep Cluster, Deep Autoencoder. 
\end{IEEEkeywords}

\section{Introduction}

Quantum communication offers security based on quantum principles, with \ac{QKD} creating cryptographic keys that are information-theoretically secure \cite{Scarani2009, gisin2002QuantumQKDReview}. Terrestrial QKD faces a limitation of ~100-200 km due to signal loss in optical fibers \cite{gisin2002QuantumQKDReview}. Satellite quantum communication bypasses this with minimal free-space absorption and wide global reach, allowing secure channels over intercontinental ranges.

The achievable rate (maximum secure key bits per unit time) is crucial for satellite \ac{QKD} viability. Estimating this rate is difficult due to complex noise (atmospheric turbulence, background radiation, pointing errors, detector flaws) degrading quantum state fidelity. The stochastic, variable nature of satellite channels makes traditional analysis insufficient for capturing non-Gaussian noise and complex correlations. Advances like the Micius satellite's over 1 kbps secure key rates across 1200 km show the feasibility of space-based quantum networks \cite{liao2017satqkd}. Further developments include Gaussian approximations for atmospheric effects, semi-analytical secret key rate methods, and machine learning for parameter estimation. Current models, however, rest on simplified noise assumptions or need high computational power. 

Unlike traditional clustering approaches \cite{mouliMECOM2024}, which operate directly on high-dimensional raw features and assume Gaussian-distributed noise, the \ac{DAGMM} \cite{zong2018DAGMM} leverages deep autoencoders to extract nonlinear latent representations that capture the complex noise correlations and non-Gaussian statistics inherent in satellite quantum channels. This learned feature space enables more accurate clustering of channel states and tighter capacity bounds, particularly in low \ac{SNR} regimes where conventional \ac{GMM}s fail to discriminate between subtle variations in hybrid noise patterns.

ML-based \ac{GMM}-EM framework lacks the capacity to capture complex, nonlinear, and Poissonian quantum noise characteristics, making it suboptimal for enhancing channel achievable rates \cite{ Mouli2025ML}. It struggles with local optima, poor feature representation, and fixed Gaussian assumptions. In contrast, the \ac{DAGMM} framework leverages deep autoencoders to learn compact, nonlinear latent features, enabling robust modeling of diverse quantum noise patterns \cite{zong2018DAGMM}. This leads to superior noise mitigation, better generalization across dynamic quantum channels, and significant achievable rate improvements.

This work builds upon these foundations by introducing a \ac{DAGMM} framework that learns compact representations of high-dimensional channel states, enabling more accurate estimation of achievable rates in satellite quantum communication systems under realistic hybrid noise conditions. We demonstrate that while classical \ac{GMM} performs reasonably well under idealized channel assumptions, its effectiveness diminishes in the presence of structured or high-dimensional noise. In contrast, the \ac{DAGMM} framework consistently yields more accurate capacity estimates, particularly in low signal-to-noise scenarios and non-Gaussian noise environments. These findings underscore the importance of combining representation learning with probabilistic modeling for scalable and accurate capacity estimation in space-based \ac{QKD} systems.

The contributions of this work are threefold:
\begin{itemize}
      \item We formulate a simulation-based framework for capacity estimation in satellite quantum channels, utilising both classical and deep learning-based clustering techniques.

      \item We provide a rigorous comparative evaluation between \ac{GMM} and \ac{DAGMM} models under varying noise and system conditions.

      \item We analyze the interpretability, scalability, and robustness of each model in the context of satellite \ac{QKD}, offering practical insights for real-world deployment.
\end{itemize}

\textit{Notation,} In this work work, we use $tr(\cdot)$ and $\mathbf{T(\cdot)}$ to denote the trace of a matrix and {trace-preserving} map, respectively. For a matrix $\mathbf{A}$, $\mathbf{A}^\dag$ and ${\mathbf{A}}^t$ represent the adjoint and transpose, respectively. We denote the complex conjugate of a vector $\boldsymbol{\nu}$ by using $\boldsymbol{\nu}^*$, and the tensor product is denoted by $\otimes$. We represent the Gaussian density by $\mathcal{N}$. For a quantum state $\psi$, ket and bra are denoted by $\ket{\boldsymbol{\psi}}$ and $\bra{\boldsymbol{\psi}}$, respectively. 

\section{System Model}


A qubit is a superposition $\ket{\boldsymbol{\psi}} = \boldsymbol{\alpha}\ket{\boldsymbol{0}} + \boldsymbol{\beta}\ket{\boldsymbol{1}}$, with $\boldsymbol{\alpha}$ and $\boldsymbol{\beta}$ as complex numbers satisfying $|\boldsymbol{\alpha}|^2 + |\boldsymbol{\beta}|^2 = 1$; its pure state density matrix is $\boldsymbol{\rho} = \ket{\boldsymbol{\psi}}\bra{\boldsymbol{\psi}} = \begin{bmatrix} |\boldsymbol{\alpha}|^2 & \boldsymbol{\alpha}^*\boldsymbol{\beta} \\ \boldsymbol{\alpha\beta}^* & |\boldsymbol{\beta}|^2 \end{bmatrix}$. Mixed states are represented by $\boldsymbol{\rho} = \sum_i p_i \ket{\boldsymbol{\psi}_i}\bra{\boldsymbol{\psi}_i}$ with $\sum_i p_i = 1$ \cite{nielsen_chuang_2010}. In quantum mechanics, a pure-state qubit maps to a Bloch sphere point $(\theta, \phi)$ via function $\zeta(\theta, \phi)$, mainly affected by noise in $\theta$ while $\phi$ stays almost constant, transforming as $(\theta, \phi) \mapsto (\tilde{\theta}, \phi \pm \delta)$. Mixed states introduce radial coordinate $r$, yielding $(\theta, \phi, r)$ with $\tilde{\zeta}(\theta, \phi, r)$. Noise causes minor shifts in $\phi$ and $r$ (variations $\delta_1$ and $\delta_2$ treated similarly), approximating qubit motion as rotation along a circular or spiral path. Scalar $\theta$ describes noise-influenced behavior using one-dimensional randomness, capturing key probabilistic features and reducing computational demands \cite{mouli2024}.

In quantum optics, quadrature operators for position $\hat{\boldsymbol{q}} = \frac{1}{\sqrt{2}}(\hat{\boldsymbol{a}} + \hat{\boldsymbol{a}}^\dagger)$ and momentum $\hat{\boldsymbol{p}} = \frac{1}{\sqrt{2}i}(\hat{\boldsymbol{a}} - \hat{\boldsymbol{a}}^\dagger)$ satisfy the canonical commutation relation $[\hat{\boldsymbol{q}}, \hat{\boldsymbol{p}}] = i$, forming the Hamiltonian of a single-mode quantum harmonic oscillator $\hat{\mathbf{H}} = \frac{1}{2}(\hat{\boldsymbol{p}}^2 + \hat{\boldsymbol{q}}^2)$. A quantum channel $\mathcal{N}$ is a CPTP map using Kraus operators $E_k$ to model noise and decoherence \cite{wilde2013quantum}. It can also be seen as a unital map $\mathbf{N}: A_1 \to A$ that transforms input $\boldsymbol{\rho}$ to output $\boldsymbol{\rho} \circ \mathbf{N}$, with quantum communication occurring through encoding, channel action, and decoding $C_1 \xrightarrow{\gamma'} A_1 \xrightarrow{a'} A \xrightarrow{\delta'} C$ \cite{ohya2004quantum}. Mutual entropy $I(\boldsymbol{\rho}; \mathbf{N})$ measures input-output correlation and efficiency. Quantum channels alter density matrices via environment-assisted unitary operations $\boldsymbol{\rho} \mapsto \operatorname{tr}_E[\boldsymbol{U}(\boldsymbol{\rho} \otimes \boldsymbol{\rho}_E)\boldsymbol{U}^\dagger]$, using a unitary $\boldsymbol{U}$ and environment state $\boldsymbol{\rho}_E$. Gaussian channels change covariance matrices $\boldsymbol{\rho} \mapsto \boldsymbol{A}^t\boldsymbol{\rho}\boldsymbol{A} + \boldsymbol{Z}$, with a real matrix $\boldsymbol{A}$ controlling amplification, attenuation, and rotation, while $\boldsymbol{Z}$ includes quantum and classical noise \cite{cerf2007quantum}.

Alice generates \( 2N \) Gaussian random variables, \( X_1, X_2, \ldots, X_{2N} \), with variance \(\sigma_X^2\), to represent \( N \) coherent states \(\ket{\boldsymbol{\alpha}_n} = X_{2n-1} + iX_{2n}\). State amplitudes and phases are encoded using these variables. Bob measures the quadratures \(X_{2n-1}\) and \(X_{2n}\) through homodyne detection, obtaining \(Y_{2n-1}\) and \(Y_{2n}\), forming a key \(\boldsymbol{Y} = (Y_1, Y_2, \ldots, Y_{2N})\). This is effective with low transmission and reverse reconciliation. \ac{HQN}, $Z$, is modeled as the convolution of quantum Poissonian noise ($Z^{(1)}$) and classical \ac{AWGN} ($Z^{(2)}$) \cite{mouli2024, Mouli2024Asymp_QKD_SatComm, Mouli2024Finite_size_QKD_QComm}. Specifically, $Z = Z^{(1)} + Z^{(2)}$, where $Z^{(1)}$ follows a Poisson distribution \cite{fox2006quantum} and $Z^{(2)}$ is Gaussian-distributed with mean $\mu_{Z^{(2)}}$ and variance $\sigma_{Z^{(2)}}^2$. The \ac{p.d.f.} of \ac{HQN} is represented as a \ac{GMM} \cite{mouli2024, Mouli2024Asymp_QKD_SatComm,  mouliMECOM2024,  Mouli2025ML},
\begin{multline}
    f_{Z}(z) = \sum_{i=0}^{R} \frac{e^{-\lambda} \lambda^i}{i!} \frac{1}{\sigma_{Z^{(2)}}\sqrt{2\pi}} e^{-\frac{1}{2}\left(\frac{z-r-\mu_{Z^{(2)}}}{\sigma_{Z^{(2)}}}\right)^2}
    \\= \sum_{i=0}^{R} w_i \, \mathcal{N} \left(z; \mu_i^{(z)}, \, {\sigma_i^{(z)}}^2\right)
\end{multline}
where $w_i = \frac{e^{-\lambda}\lambda^i}{i!}$ are the mixture weights and $R$ is the number of Gaussian clusters. For higher dimensions, the multivariate \ac{p.d.f.} of \ac{HQN}, $\mathbf{z} \in \mathbb{R}^D$, is given by,

\begin{equation}
    f_{\boldsymbol{Z}}(\mathbf{z}) = \sum_{i=0}^{R} w_{i}.\mathcal{N} \Big(\mathbf{z};\boldsymbol{\mu}_{i}^{(z)},\,{\boldsymbol{\Sigma}_{i}^{(z)}}\Big),
    \label{eq approx pdf of HQN}
\end{equation}
where 
\begin{multline}
      \mathcal{N}(\mathbf{z};\boldsymbol{\mu}_{i}^{(z)},\boldsymbol{\Sigma}_{i}^{(z)}) =\\ \frac{1}{(2\pi)^{D/2}|\boldsymbol{\Sigma}_{i}^{(z)}|^{1/2}}e^{-\frac{1}{2}(\mathbf{z}-\boldsymbol{\mu}_{i}^{(z)})^{T}\{\boldsymbol{\Sigma}_{i}^{(z)}\}^{-1}(\mathbf{z}-\boldsymbol{\mu}_{i}^{(z)})},
\end{multline} with $\boldsymbol{\mu}_{i}^{(z)}$ as the mean vector and $\boldsymbol{{\Sigma}}_{i}^{(z)}$ as the covariance matrix. This formulation effectively represents HQN as a weighted sum of multivariate Gaussian components.
From this generalized description of a quantum channel, we can drive towards the atmospheric \ac{FSO} satellite quantum channel model represented by the following equation \cite{Mouli2024Asymp_QKD_SatComm, dequal2021_SAT_CV_QKD},
\begin{equation}
    Y=TX +Z,
    \label{eq atmospheric quantum channel}
\end{equation} 
where the quantum channel transmitted signal is given by random variables $X$, representing Alice's inputs, and the received signal $Y$ at Bob's side, $Z$ is the \ac{HQN} given by \eqref{eq approx pdf of HQN}, where $T$ is the overall transmission coefficient for the single use of the quantum channel. The focus should be on the transmission coefficient $T$ (with the transmission efficiency $\tau= T^2$).  This model captures the complexity of satellite quantum communication, underscoring the importance of accounting for various noise sources and channel imperfections in ensuring secure communication.

The \ac{p.d.f.} of received signal $Y$ can be evaluated as convolution of transmitted signal and the \ac{HQN}, given as, 
\begin{multline}
    f_{Y}(y) \\
        = \sum_{j=0}^{R}
             \frac{e^{-\lambda}\lambda^j}{j!}\frac{1}{\sqrt{2\pi(T^2\sigma_{X}^2 +\sigma_{Z^{(2)}}^2)}}
             e^{\Biggl[{-\frac{\big(y-j-T\mu_{X}-\mu_{Z^{(2)}}\big)^2}{2\big(\sigma_{X}^2 +T^2\sigma_{Z^{(2)}}^2\big)}}\Biggl]}\\
        =\sum_{r=0}^{R}\pi_{r}^{(Y)}\mathcal{N} \Big(y,\mu_{r}^{(Y)}, \,{\sigma_{r}^{(Y)}}^{2}\Big),
  \label{eq final p.d.f. of Y2}
\end{multline} 
where \( \pi_r^{(Y)} = \frac{e^{-\lambda} \lambda^r}{r!} \), $\sum_{r=0}^{R} u_{r}^{(Y)} \approx 1$ for large $R$, \( \mu_r^{(Y)} = T\mu_{X} + \mu_{Z^{(2)}} + r \), and \( \sigma_r^{(Y)} = \sqrt{T^2 \sigma_{X}^2 + \sigma_{Z^{(2)}}^2} \). The corresponding multivariate Gaussian mixture for received sianl \ac{p.d.f.} is given by \cite{Mouli2024Asymp_QKD_SatComm}
\begin{equation}
   f_{\boldsymbol{Y}}(\boldsymbol{y}) = \sum_{r=0}^{R} \pi_{r}^{(\boldsymbol{Y})}\mathcal{N} \Big(\boldsymbol{y};\boldsymbol{\mu}_{r}^{(\boldsymbol{Y})},\, \boldsymbol{\Sigma}_{r}^{(\boldsymbol{Y})}\Big), 
    \label{eq approx. p.d.f. of received signal vector}   
\end{equation}
where $\boldsymbol{\mu}_{r}$ is the mean vector, and $\boldsymbol{\Sigma}_{r}^{(\boldsymbol{Y})}$ is the covariance matrix of the corresponding Gaussian density $\mathcal{N}$. 

\subsection{Hybrid Quantum Noisy Channel optimization with Autoencoder}  

An autoencoder is a neural network designed to learn a compressed representation of data \cite{Autoencoder2006GEHinton}. It consists of two main parts: an encoder and a decoder. The encoder maps the high-dimensional input data \( \boldsymbol{x} \) to a lower-dimensional latent representation \( \hat{\boldsymbol{x}} \), capturing the essential features of the data. The decoder then reconstructs the input data from this latent representation \( \hat{\boldsymbol{x}} \). For a deep autoencoder that consists of an encoder and a decoder, let us denote $\boldsymbol{\theta}=({\boldsymbol{\theta}}_1, {\boldsymbol{\theta}}_2)$ as the total network parameters, where $\boldsymbol{\theta}_1$ indicates the encoder parameters and ${\boldsymbol{\theta}}_2$ means the decoder parameters.


The encoder can be represented mathematically as,
\begin{equation} \hat{\boldsymbol{x}} = f_{\text{encoder} (\boldsymbol{\theta}_1)}(\boldsymbol{x}) \end{equation}
where \( f_{\text{encoder}} \) consists of several layers,
\( h^l = \sigma(W^l h^{l-1} + b^l) \),
with \( h^0 = \boldsymbol{x} \), \( W^l \) and \( b^l \) being the weight matrix and bias vector of the \( l \)-th layer, and \( \sigma \) being an activation function.

The decoder is similarly defined as,
\begin{equation} \boldsymbol{y} = f_{\text{decoder} (\boldsymbol{\theta}_2)}(\hat{\boldsymbol{x}}) \end{equation}
where \( f_{\text{decoder}} \) consists of layers, \( \hat{h}^l = \sigma(\hat{W}^l \hat{h}^{l-1} + \hat{b}^l) \), with \( \hat{h}^0 = \hat{\boldsymbol{x}} \), \( \hat{W}^l \) and \( \hat{b}^l \) being the weights and biases of the \( l \)-th layer in the decoder.

The objective of training the autoencoder is to minimize the reconstruction error between the input \( \boldsymbol{x} \) and the reconstructed output \( \boldsymbol{y} \). The \ac{MSE} loss function is used,
\[ \mathcal{L}_{\text{MSE}} = \frac{1}{R} || \boldsymbol{x}_i - \boldsymbol{y}_i ||^2 \]
Training involves updating the weights using gradient-based optimization methods with gradients computed through backpropagation.

Once the autoencoder is trained, the representations \( y \) are obtained through \ac{GMM},  since the data is generated from a mixture of Gaussian distributions,

\begin{equation}
    f_{\boldsymbol{Y}}(\boldsymbol{y}_{i}) = \sum_{r=1}^{R} \pi_r \mathcal{N}(\boldsymbol{y}_{i}|\boldsymbol{\mu}_r, \boldsymbol{\Sigma}_r),
\end{equation}
where \( R \) is the number of Gaussian components (= number of effective clusters based on Poisson weightage), \( \pi_r \) are the mixing coefficients,  \( \sum \pi_r \approx 1\) and \( \mathcal{N}(\boldsymbol{y}_{i}|\boldsymbol{\mu}_r, \boldsymbol{\Sigma}_r) \) is the Gaussian distribution with mean \(\boldsymbol{\mu}_r \) and covariance \( \boldsymbol{\Sigma}_r \).

\subsection{Deep Cluster Gaussian Mixture Model in Satellite Quantum Communication}

Inspired by \ac{DNN}, we define a \ac{DGMM} as a network of latent variable layers, each following a Gaussian mixture, forming nested linear models that provide flexible nonlinear modeling.

For \( h \) layers and observed data \( \boldsymbol{y} \) , the model per observation \( i \) is:
\begin{multline}
\boldsymbol{y}_i = T_{s1}^{(1)} \boldsymbol{x}_i^{(1)} + \boldsymbol{z}_i^{(1)}  \quad \text{with probability} \, \pi_{s1}^{(1)}, \, s1 = 1, \ldots, r_1
\\ 
\boldsymbol{x}_i^{(1)} =  T_{s2}^{(2)} \boldsymbol{x}_i^{(2)} + \boldsymbol{z}_i^{(2)} \, \text{with probability} \, \pi_{s2}^{(2)}, \, s2 = 1, \ldots, r_2
\\
\vdots
\\
\boldsymbol{x}_i^{(h-1)} = T_{sh}^{(h)} \boldsymbol{x}_i^{(h)} + \boldsymbol{z}_i^{(h)} \, \text{with probability} \, \pi_{sh}^{(h)}, \, sh = 1, \ldots, r_h
\end{multline}
where \( \boldsymbol{x}_i^{(h)} \) follows a Gaussian with zero mean and unity variance, and \( \boldsymbol{z}_i^{(1)}, \ldots, \boldsymbol{z}_i^{(h)} \) are \ac{HQN} following \ac{GMM} \eqref{eq approx pdf of HQN}, $ \forall i = 1, \ldots, 2N$.  \( T_{s1}^{(1)}, \ldots, T_{sh}^{(h)} \) are the transmission coefficients of each quantum link, with transmitted signal data \( \boldsymbol{x} \) independent of  \ac{HQN} data \( \boldsymbol{z} \). Thus, each layer's conditional distribution given latent variables is a multivariate Gaussian mixture.

Let $\Omega$ be all network paths, where path $l = (l1, \cdots , lh)$ has probability $\pi_l$ with $\sum_{l \in \Omega} \pi_l =1$. The \ac{DGMM} is given by, 
$
f (\hat{\boldsymbol{x}};\Theta ) =\sum_{l \in \Omega} \pi_{l} \mathcal{N}(\hat{\boldsymbol{x}}; \boldsymbol{\mu}_{l}, \boldsymbol{\Sigma}_{l} ).
$ 

The algorithm alternates between two steps, consisting of maximizing (M-step) and calculating the conditional expectation (E-step) of the complete-data log-likelihood function given the observed data, evaluated at a given set of parameters \( \Theta'(l, \pi, \mu, \sigma) \), 
\(
\mathbb{E}_{x_i^{(1)}, \ldots, x_i^{(l)}, t | y; \Theta'} \left[ \ln \mathcal{L}(\Theta) \right]
\), and $ \ln \mathcal{L}=  \ln p(\hat{\boldsymbol{x}}|\boldsymbol{\pi},\boldsymbol{\mu}, \boldsymbol{\Sigma} )= \sum_{i=1}^{N} \ln \biggl[ \sum_{r=1}^{R} \pi_{r}\mathcal{N}(\hat{\boldsymbol{x}}_r|\boldsymbol{\mu}_r, \boldsymbol{\Sigma}_r) \biggl]$ . Unlike classical \ac{GMM} (which computes only allocation posteriors for $l$), \ac{DGMM} requires multivariate posteriors, making EM slow and unscalable.

The objective function combines the reconstruction error, the log-likelihood of the quantum data $Y$ under the \ac{GMM}, 
\begin{equation}
    \begin{split}
   & min : \frac{1}{N}\sum_{i=1}^N \bigg( \| \boldsymbol{x}_i - \boldsymbol{y}_i \|^2 - \tilde{\lambda} \ln \sum_{r=1}^R \pi_r p(\hat{\boldsymbol{x}}_i |\boldsymbol{ \mu}_r, \boldsymbol{\Sigma}_r)\bigg)  \\
    & s.t. \, \, \,  \hat{\boldsymbol{x}}_i = f_{\boldsymbol{\theta}}(\boldsymbol{x}_i) , \, \, i= 1, ..., R.
        \label{}
   \end{split}
\end{equation} 
The objective function combines the reconstruction error, the log-likelihood of the data under the \ac{GMM}, and a penalty term for the constraint. The augmented Lagrangian \( \mathcal{L}_{\tilde{\rho}} \) is
\begin{equation}
    \begin{split}
        & \mathcal{L}_{\tilde{\rho}} (\boldsymbol{\theta}, \boldsymbol{Y}, \boldsymbol{U}, \boldsymbol{\mu}, \boldsymbol{\Sigma}, \boldsymbol{\pi}) \\
        & = \sum_{i=1}^N \bigg( \|  \boldsymbol{x}_i -  \boldsymbol{y}_i \|^2 - \tilde{\lambda} \ln \sum_{r=1}^R \pi_k p(\hat{\boldsymbol{x}}_i |  \boldsymbol{\mu}_r,  \boldsymbol{\Sigma}_r)\\
        & + \tilde{\rho} \|  \hat{\boldsymbol{x}}_i - f_{ \boldsymbol{\theta}}( \boldsymbol{x}_i) +  \boldsymbol{z}_i \|^2 \bigg) ,
        \label{}
    \end{split}
\end{equation} 
where \(\|  \boldsymbol{x}_i -  \boldsymbol{y}_i \|^2\) is the reconstruction error, \(\ln \sum_{r=1}^R \pi_r p(\hat{\boldsymbol{x}}_i | \boldsymbol{\mu}_r, \boldsymbol{\Sigma}_r)\) is the log-likelihood of \( \hat{\boldsymbol{x}}_i \) under the \ac{GMM}, $\boldsymbol{U}= \hat{\boldsymbol{x}}_i - f_{\theta}(\boldsymbol{x}_i)$ is the scaled dual variable reciprocal of $\tilde{\lambda}$, \(\| \hat{\boldsymbol{x}}_i - f_{{\boldsymbol{\theta}}}({\boldsymbol{x}}_i) + \boldsymbol{z}_i \|^2\) is the term for associated with \ac{HQN}, \(\tilde{\lambda }\) is the Lagrange multipliers and \( \tilde{\rho} \) as the penalty parameter.
Using the probability \(p(t_{ir}=1|\hat{\boldsymbol{x}}_{ir})\), as responsibility $\gamma(t_{ir},\boldsymbol{x}_{ir})$ given by, 
\begin{equation}
    \gamma(t_{ir},\boldsymbol{x}_{ir}) = \frac{\pi_r p(\hat{\boldsymbol{x}}_i | \boldsymbol{\mu}_r, \boldsymbol{\Sigma}_r)}{\sum_{j=1}^R \pi_j p(\hat{\boldsymbol{x}}_i | \boldsymbol{\mu}_j, \boldsymbol{\Sigma}_j)}. \label{}
\end{equation} 
\textit{Minimizing the Augmented Lagrangian :}
To find the minimum, we take the derivative of \( \mathcal{L}_{\tilde{\rho}} \) with respect to \( \hat{\boldsymbol{x}}_i \) and set it to zero,
\begin{multline}
    \frac{\partial \mathcal{L}_{\tilde{\rho}}}{\partial \hat{\boldsymbol{x}}_i}
 = - \lambda \sum_{r=1}^R \gamma(\boldsymbol{x}_{ir}) \frac{\partial}{\partial \hat{\boldsymbol{x}}_i} \biggl(\ln p(\hat{\boldsymbol{x}}_i | \boldsymbol{\mu}_k, \boldsymbol{\Sigma}_k) \\  + 2{\tilde{\rho}} (\hat{\boldsymbol{x}}_i - f_{\theta_1}(\boldsymbol{x}_i) + \boldsymbol{z}_i) \biggl)
 = 0.
 \label{}
\end{multline}

Solving for \( \boldsymbol{y}_i \), 
and hence the update equation is given by,
\begin{equation}
    \begin{split}
        & \boldsymbol{y}_i^{'}
        = \bigg( \tilde{\lambda} \sum_{r=1}^R \gamma(\boldsymbol{x}_{ir}) \boldsymbol{\Sigma}_r^{-1} + {\tilde{\rho}} I \bigg)^{-1}\\ & \bigg( \sum_{r=1}^R \gamma(\boldsymbol{x}_{ir}) \boldsymbol{\Sigma}_r^{-1} \mu_r + {\tilde{\rho}} (f_{\boldsymbol{\theta}}(\boldsymbol{x}_i) - \boldsymbol{z}_i) \bigg).
         \label{}
    \end{split}
\end{equation}
This equation combines the responsibilities, the means, and covariances of the Gaussian components, as well as the penalty term to ensure the constraints are satisfied. Other parameters are given by,
\begin{align}
\bm{\mu}_r &= \frac{1}{N_k} \sum_{i=1}^{N} \gamma_i(t_{ir}) \hat{\boldsymbol{x}}_i \notag \\
\bm{\Sigma}_r^{'} &= \frac{1}{N_r} \sum_{i=1}^{N} \gamma_i(t_{ir}) (\hat{\boldsymbol{x}}_i - \bm{\mu}_r^{'})(\hat{\boldsymbol{x}}_i - \bm{\mu}_r^{'})^\top \notag \\
\pi_r^{'} &= \frac{N_r}{N}, \quad \text{where } N_r = \sum_{i=1}^{N} \gamma_i(t_{ir}) \notag \\
\end{align}

\section{Numerical Analysis}

Synthetic quantum satellite communication data is generated and visualized in 3D before clustering. The autoencoder trains over multiple epochs with decreasing reconstruction loss while tracking \ac{GMM} log-likelihood convergence, after which latent representations are clustered and visualized.

\begin{figure*}[!]
    \subfloat[\centering ]{{\includegraphics[width=2.15in]{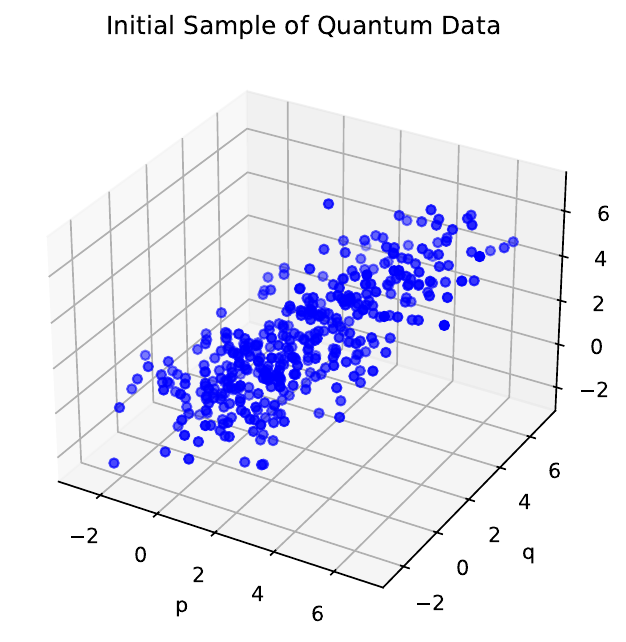} }}%
    \qquad
    \subfloat[\centering ]{{\includegraphics[width=2.15in]{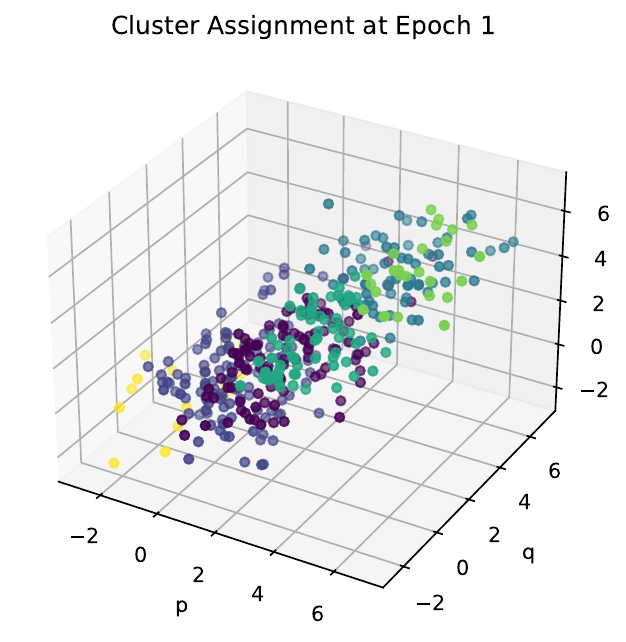} }}
    \qquad
    \subfloat[\centering ]{{\includegraphics[width=2.15in]{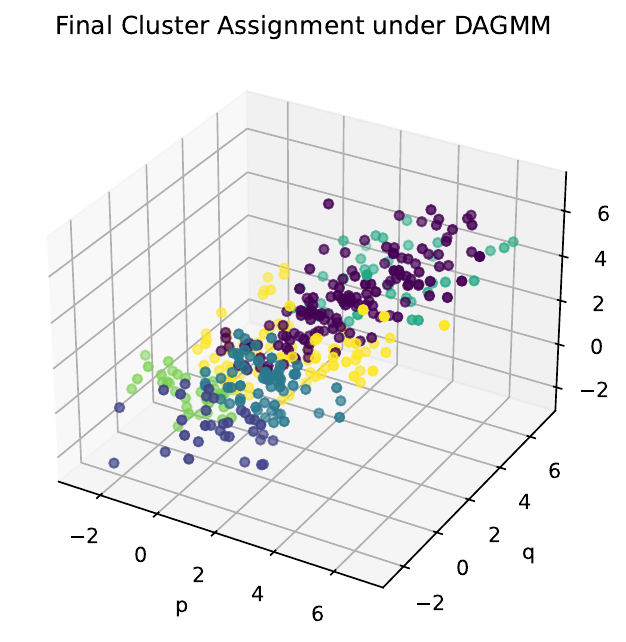} }}
    \caption{(a) Initial distribution of simulated quantum satellite communication data before clustering, visualized in three-dimensional feature space, (b) Cluster assignments at the first epoch using \ac{DAGMM}, showing the initial latent structure during training, (c) Final cluster assignment obtained from \ac{DAGMM} after convergence, reflecting well-separated clusters in the latent space.}%
    \label{fig 1}%
\end{figure*}
\begin{figure*}[!]
    \subfloat[\centering ]{{\includegraphics[width=2.15in]{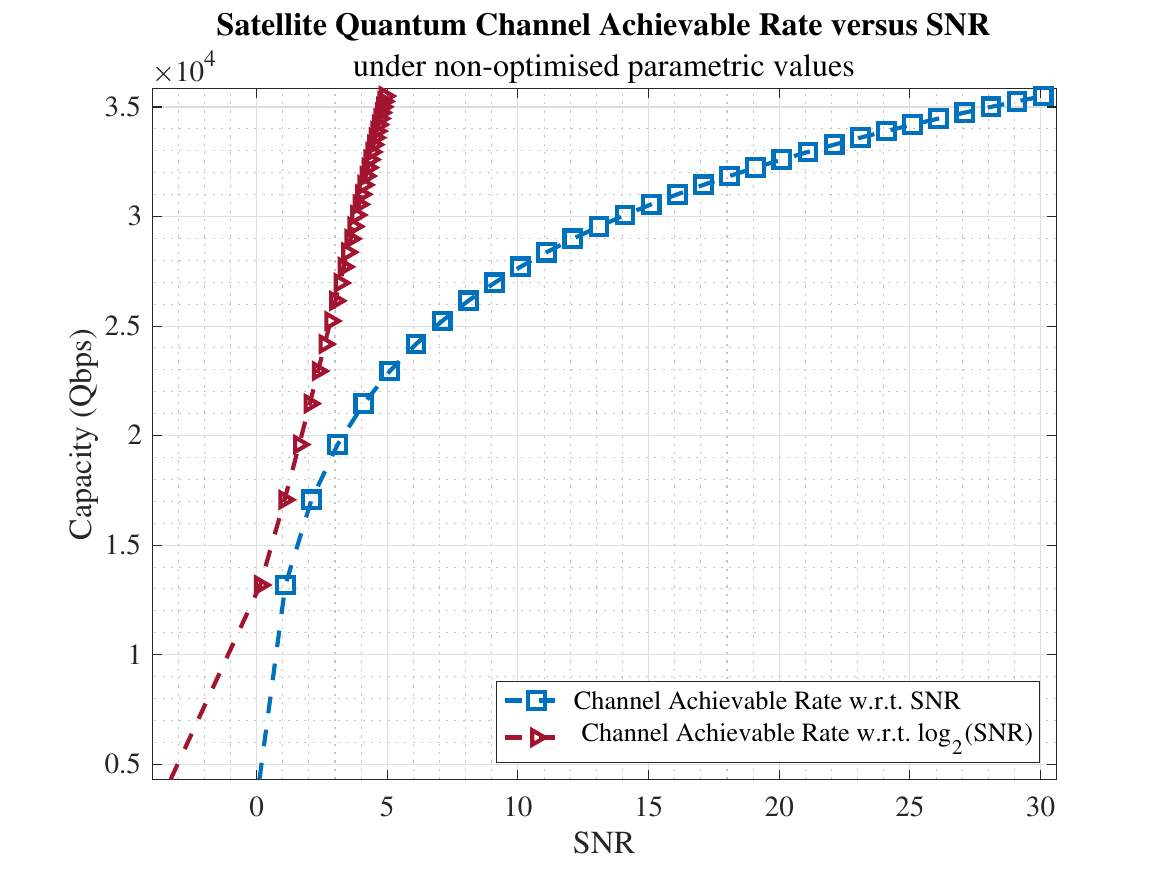} }}%
    \qquad
    \subfloat[\centering ]{{\includegraphics[width=2.15in]{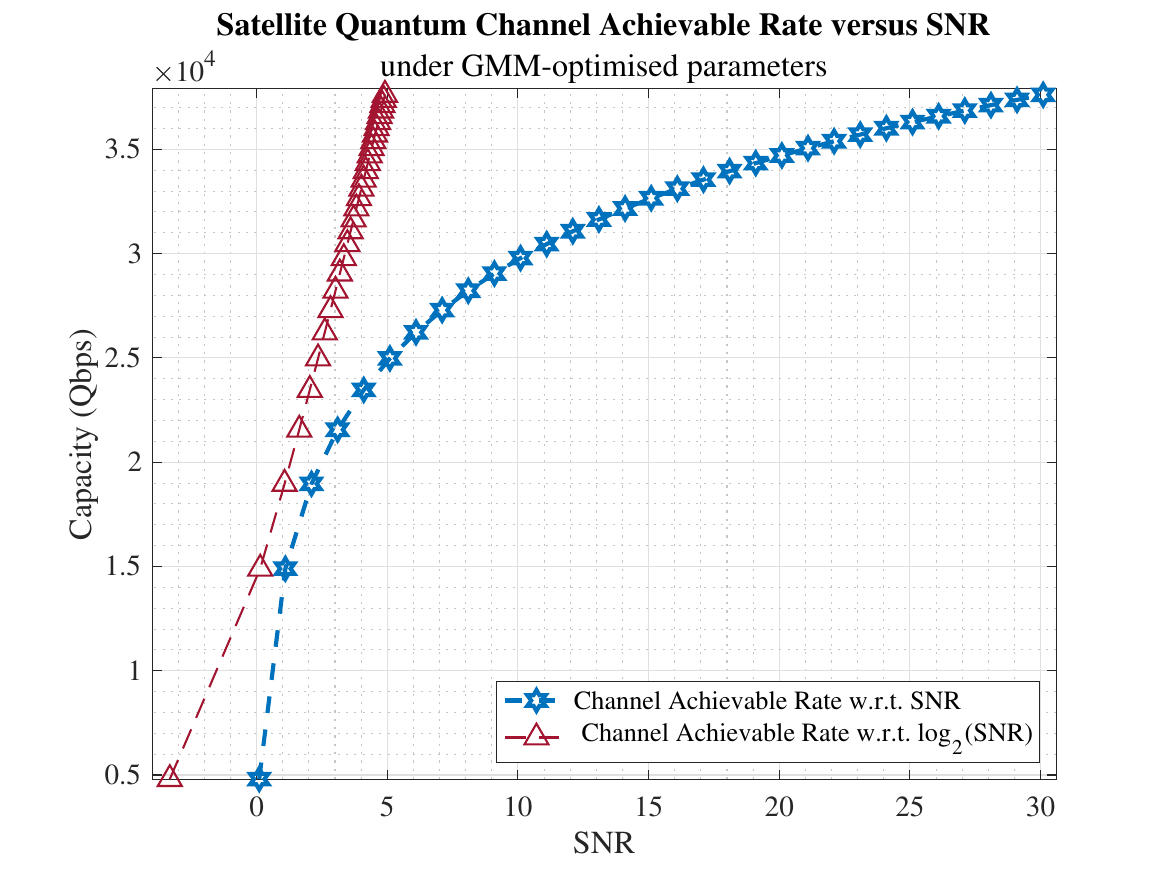} }}
    \qquad
    \subfloat[\centering ]{{\includegraphics[width=2.15in]{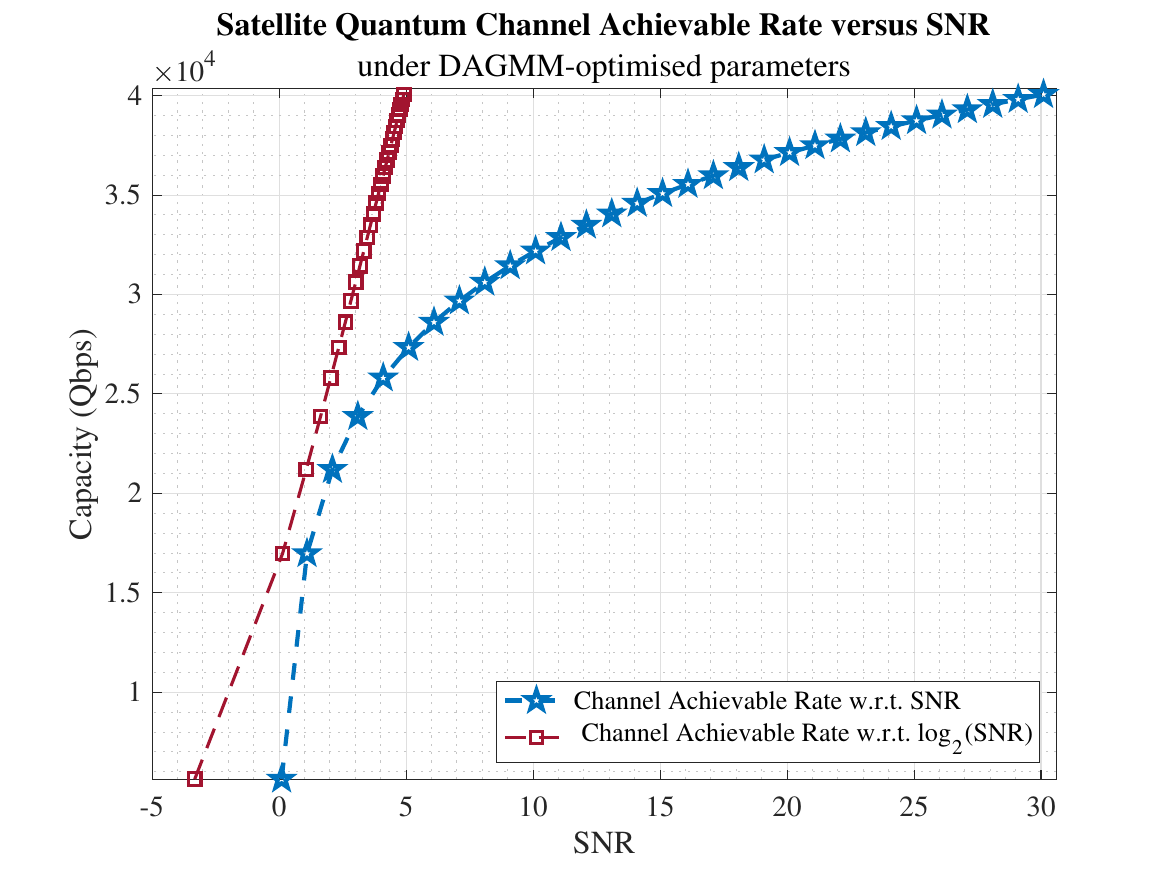} }}
    \caption{(a) Achievable quantum channel capacity versus \ac{SNR} using default parameters, without \ac{GMM}-based optimization, (b) Channel capacity as a function of \ac{SNR} using \ac{GMM}-optimized parameters, (c) Quantum channel capacity versus \ac{SNR} using \ac{DAGMM} optimization. }%
    \label{fig 2}%
\end{figure*}
\subsection{Parameter Description for \ac{QKD} over Satellite Quantum Channel}

The \ac{QKD} satellite channel parameters include: Poisson parameter $\lambda=3$ (average photons/pulse), 6 Gaussian clusters, reconciliation efficiency $\beta=0.95$, electronic noise \(\nu_{ele}= 0.041 \), detection efficiency \( \eta_{det} =0.606\), receiver telescope aperture $\epsilon=0.005$, satellite altitude $h=20,000$km, photon temporal width = $3$ ns, and beam waist \( w_0 = 5\)  cm. Optical link parameters:  transmitter aperture \( a_S = 10\)  cm, receiver aperture \( a_r = 30\)  cm, wavelength \( w = 800\)  nm. Gravitational parameters: \( G = 6.674 × 10^{-11}\) Nm²/kg², Earth's mass \( M_E= 5.972 × 10^{24} \)  kg.

\subsection{Clustering of Satellite Quantum Data}

Fig. \ref{fig 1} (a) presents the raw, high-dimensional data generated under a realistic satellite quantum channel model. It serves as the baseline for evaluating clustering algorithms. The unstructured distribution highlights the challenge of distinguishing signal regimes without the application of any representation learning or clustering.
Fig. \ref{fig 1} (b) presents the clustering of \ac{HQN} data set at epoch 1; the latent representations are poorly separated, indicating that the encoder has not yet learned meaningful features. This early snapshot highlights the need for iterative training in deep models to uncover useful clustering structure in quantum channel data.
In Fig. \ref{fig 1}, (c), the result illustrates that the deep autoencoder has successfully learned a compressed representation that reveals distinct cluster structures in the satellite quantum data. The improved separation compared to Fig. 2 supports the use of deep clustering in high-noise, high-dimensional scenarios for quantum signal analysis.

\subsection{Optimization of Satellite Quantum Channel}

Fig. \ref{fig 2} (a) represents the baseline result of achievable rate estimation of the satellite \ac{HQN} channel. This result demonstrates the capacity performance when physical and signal processing parameters are not tuned. The observed achievable rate is lower, particularly at high \ac{SNR}, emphasizing the need for optimized channel modeling techniques.
Fig. \ref{fig 2} (b) presents the classical \ac{GMM}-based achievable rate estimation of the satellite \ac{HQN} channel. \ac{GMM} improves achievable rate estimation over the non-optimized baseline, particularly in mid-to-high \ac{SNR} regimes. However, the performance remains limited by its reliance on raw feature space, which makes it less effective in capturing deep, non-linear channel dynamics. In Fig. \ref{fig 2} (c), the \ac{DAGMM}-based optimisation of the satellite \ac{HQN} channel helps in achieving the highest achievable rate among all models, particularly in low \ac{SNR} regimes. This underscores its ability to effectively model \ac{HQN} and extract meaningful latent features, thereby providing tighter and more robust capacity bounds for satellite \ac{QKD}.

\section{Conclusion}

In this work, we demonstrated that classical \ac{GMM}, while interpretable and computationally efficient, struggles to accurately model high-dimensional, non-linearly structured quantum channel data. \ac{DAGMM} addresses this limitation by learning latent representations through deep autoencoders, which significantly improve clustering visualization by separating complex noise components in reduced feature space. The enhanced cluster fidelity enables more precise identification of effective transmission states and noise regimes. As a result, \ac{DAGMM} provides tighter and more reliable estimates of quantum channel capacity compared to traditional \ac{GMM}. This establishes \ac{DAGMM} as a more robust and scalable solution for performance optimization in satellite-based \ac{QKD} systems.
These findings establish deep representation learning as a powerful quantum channel modeling tool for satellite quantum links, with future work targeting dynamic channel environments, time-varying parameters, and hybrid architectures combining classical physical models with deep neural encoders for real-time \ac{QKD} performance prediction.

\begin{acronym}
\acro{DA}{Deep Autoencoder}
\acro{DCGMM}{Deep Cluster Gaussian Mixture Model}
\acro{DAGMM}{Deep Autoencoder Gaussian Mixture Model }
\acro{HQC}{hybrid quantum channel}
\acro{HQNM}{hybrid quantum noise model}
\acro{HQN}{hybrid quantum noise}
     \acro{RSA}{Rivest, Shamir and Adelman algorithm}
     \acro{ECC}{elliptic curve cryptography}
     \acro{LOQC}{linear optical quantum computing}   
     \acro{CPTP}{completely positive trace-preserving}
     \acro{MAP}{maximum a posteriori}
    \acro{ML}{machine learning}
    \acro{DL}{deep learning}
    \acro{RL}{reinforcement learning}
    \acro{SVM}{support vector machines}
    \acro{PCA}{principal component cnalysis}
    \acro{KF}{Kalman Filter}
    \acro{DBSCAN}{density-based spatial clustering of applications with noise}
    \acro{WCSS}{within-cluster sum of squares}
    \acro{QML}{quantum machine learning}
    \acro{NN}{neural network}
    \acro{EM}{expectation-maximization}
    \acro{GM}{Gaussian mixture}
    \acro{p.d.f.}{probability density function}
    \acro{p.m.f.}{probability mass function}
    \acro{SNR}{signal-to-noise ratio}
    \acro{SPS}{single-source photon}
    \acro{GMM}{Gaussian mixture model}
    \acro{GMs}{Gaussian mixtures}
    \acro{BIC}{Bayesian information criterion }
    \acro{AIC}{Akaike information criterion}   
    \acro{AWGN}{additive-white-Gaussian noise}
    \acro{OGS}{optical ground station}
    \acro{SKR}{secret key rate}
    \acro{QKD}{quantum key distribution}
    \acro{PNS}{photon number splitting}
    \acro{CV-QKD}{continuous-cariable quantum key distribution}
    \acro{FSO}{free-space optics}
    \acro{MDI}{measure-device-independent}
    \acro{DGMM}{deep Gaussian mixture model}
    \acro{DV-QKD}{discrete-variable quantum key distribution}
    \acro{CV}{continuous variables}
    \acro{DV}{discrete variables}
    \acro{DR}{direct-reconciliation}
    \acro{RR}{reverse-reconciliation}
    \acro{CPTP}{completely positive, trace preserving}
    \acro{PM}{prepare-and-measure}
    \acro{MSE}{mean square error}
    \acro{DNN}{deep neural networks}
    \acro{PDTC}{probability distribution of transmission coefficient}
    \acro{PDTE}{probability distribution of transmission efficiency}
    \acro{S/C}{spacecraft}
    \acro{GS}{ground station}
    \acro{LDPC}{low-density parity check}

\end{acronym}










\bibliographystyle{IEEEtran}
\bibliography{IEEEabrv,paper}
%



\end{document}